\documentclass[aps,superscriptaddress,showpacs,nofootinbib]{revtex4}

\def\comment#1{}
\input{epsf}

\begin{document}

\title{
Five-Loop Critical Temperature Shift in
Weakly Interacting
Homogeneous  Bose-Einstein Condensate
}


 \author{H. Kleinert%
\thanks{Phone: 0049 30 8383034,
Email:~kleinert@physik.fu-berlin.de,~pelster@physik.fu-berlin.de,
URL: http://www.physik.fu-berlin.de/\~{}kleinert }  \\ [3mm]
{\em Institut f\"ur Theoretische Physik,}\\
{\em Freie Universit\"at Berlin, Arnimallee 14,
14195 Berlin, Germany}\\ [5mm] }

\begin{abstract}

Using variational perturbation theory,
we calculate the shift in the critical temperature $T_c$
up to five loops
to lowest order in the scattering length   $a$
and find
$ \Delta T_c/{T_c^{(0)}}\approx
(1.14\pm0.11)
\,an^{1/3}$,
where
$n$ is the particle density.
Our result is lower
than the
 latest Monte Carlo result
$ (1.32\pm0.02)\,an^{1/3}$.

\comment{Employing two different resummation schemes we find
two different results:
one $30\%$ lower than the Monte Carlo
shift
$ \Delta T_c/ T_c\approx1.32\pm0.02$,
 the other $10\%$ lower.}

\end{abstract}

\maketitle


The
effect of a small repulsive interaction
upon the critical temperature $T_c$
 of a Bose-Einstein condensate (BEC)
has been a matter of controversy
for many years, and
the various results
 are converging
only very slowly towards
a common answer.  Here we want to contribute
 to the ongoing discussion
with
a further result obtained from
a resummation of Feynman diagrams
up to five loops.

The interacting
Bose-Einstein condensate
is described by the euclidean
action
\begin{equation}
{\cal A}_E = \int_0^\beta d\tau \int d^3 x \left\{
\psi^*({\bf x},\tau)\left(
\partial _\tau -\frac{1}{2M}\nabla^2\right)
\psi ({\bf x},\tau)
-\mu \psi^* ({\bf x},\tau) \psi ({\bf x},\tau)
+  \frac{2 \pi a}{M}  \left[\psi({\bf x},\tau)\psi^*({\bf x},\tau)
\right]^2 \right\}\;,
\label{SE}
\end{equation}
where
$M$ is the mass of the bosons,
$ \beta $  the inverse temperature in natural units with $\hbar =k_B=1$,
 $a$ is the $s$-wave scattering length,
and $\mu$ the chemical potential.
The free
system
 has a
transition
temperature
\begin{equation}
T_c^{(0)} =  \frac{2 \pi}{ M} \left[\frac{n}{\zeta(3/2)}
\right]^\frac{2}{3}\;,
\label{T0}
\end{equation}
where $n$
is the particle density.
A small relative shift
of  $T_c$ with respect
to $T_c^{(0)}$ can be calculated from the general formula
\begin{equation}
\frac{ \Delta T_c}{T_c^{(0)}}
=-\frac{2}{3}\frac{ \Delta n}{n^{(0)}},
\label{@shift0}\end{equation}
where $n^{(0)}$ is the particle density in the free condensate
and $ \Delta n$ its change
at  $T_c$
caused by the small interaction.
For small $a$, this behaves like
  \cite{gordon,second}
\begin{equation}
\frac{ \Delta T_c}{T_c^{(0)}}= c_1 a
n^{1/3} + [ c_2^{\prime} \ln(a n^{1/3}) +c_2 ] a^2
n^{2/3} + {\cal O} (a^3 n).
\label{@c1}\end{equation}
where
$c'_2=-64 \pi \zeta(1/2)/3
\zeta(3/2)^{5/3}\simeq 19.7518$
can be calculated perturbatively,
whereas
$c_1$ and $c_2$
require nonperturbative
techniques
since infrared divergences
at $T_c$ make them
 basically strong-coupling results.
The standard technique
to reach this regime
 is based  on  a
resummation of perturbation expansions
using  the renormalization group
 \cite{zj,KS}, first applied in this context by Ref.~\cite{st}.
Recently, however,
it has been shown by calculating the best known critical exponent
$ \alpha $ of superfluid helium from Satellite experiments
\cite{Lipa}
that
the accuracy of strong-coupling  results
can be surpassed
by much simpler  variational perturbation theory
\cite{SC3,SCE,KS}.

Up to now, $c_2$ has been
inferred only from Monte Carlo data to be $c_2\approx75.7\pm0.4$.
In order to find
the leading coefficient $c_1$, one may take advantage
of an important simplification  due to
 the fact that $ \Delta n$
can be calculated from
 the classical limit
of the field theory,
which is governed by
the
three-dimensional action
\begin{equation}
{\cal A}_{3d} = \beta \int d^3 x \left\{
\psi_0^*({\bf x})\left(-\frac{1}{2M}\nabla^2- \mu \right )
\psi_0 ({\bf x})+ \frac{2 \pi a}{M}  \left[\psi_0^*({\bf x}) \psi_0 ({\bf x})
\right]^2 \right\}\;.
\label{3d}
\end{equation}
 This is a special case $N=2$ of the more general
O($N$)-invariant $\phi^4$ field theory
\begin{equation}
{\cal A}_{\phi}=  \int d^3x \left [ \frac {1}{2} | \nabla \phi |^2 +
\frac {1}{2} m^2
\phi^2 + \frac {u}{4!} (\phi^2)^2
\right ] \;,
\label{action2}
\end{equation}
where the $N$-component field
$\phi =  (\phi_1, \phi_2,\dots,\phi_N)$
is related
to the original
field $\psi$
for $N=2$
 by $\psi({\bf x}) =\sqrt{MT}
[\phi_1({\bf x})+i\phi_2({\bf x})]$.
The  square mass is
$m^2=-2M \mu$, and
the quartic coupling is
$u=48 \pi a MT $.
Using
this relation, the shift of the critical temperature
(\ref{@shift0})
can be found from  the formula
\begin{equation}  \!\!\!\!\!\!\!\!\!\!\!
\frac{ \Delta T_c}{T_c^{(0)}}
\approx-\frac{2}{3}\frac{MT_c^{(0)}}{n}
\left\langle { \Delta \phi^2}\right\rangle
=-\frac{4\pi}{3}\frac{(MT_c^{(0)})^2}{n}
4!\left\langle \frac{ \Delta \phi^2}{u}\right\rangle\,a
=-\frac{4\pi}{3}(2\pi)^2\frac{1}{[\zeta(3/2)]^{4/3}}
 4!
\left\langle\frac{ \Delta \phi^2}{u}\right\rangle \,an^{1/3},
\label{@}\end{equation}
corresponding in Eq.~(\ref{@c1}) to
\begin{equation} \!\!\!\!\!\!\!\!
 c_1 \approx-1103.09
\left\langle\frac{ \Delta \phi^2}{u}\right\rangle
 .
\label{@NumF}\end{equation}
\begin{figure}[b]
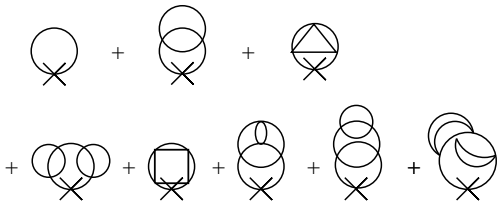

~~\\~~\\[.3cm]
~~\\~~\\
\input dia.tps
~~\\[-1cm]
\caption{Diagrams contributing to the
expectation value
 $\langle \phi^2 \rangle$.
}
\label{@FGS}\end{figure}

The three-dimensional theory is superrenormalizable and requires only
mass counterterms which shift the original bare mass $m$ to the renormalized mass $m_r$.
A calculation
of the Feynman diagrams in Fig.~\ref{@FGS}
yields  the following
 five-loop perturbation expansion  for the expectation value
$\langle \phi^2/u\rangle $ \cite{ramospr,braat}
\begin{eqnarray}
\!\!\!\!\!\!\!\!\!\!\!\!\!\!
\left\langle\frac{ \phi^2}{u}\right\rangle
&=& F\left({u}\right)\equiv -\frac{N  }{4\,\pi } \frac{m_r}{u}-
  a_2\frac{N\,
     \left( 2 + N \right) }{18\,
     {(4\pi)}^3}
\frac{u}{m_r}
 +
  a_3\frac{N\,
     \left( 16 + 10\,N + N^2 \right) }{108\,{(4\pi)}^5}
\left(\frac{u}{m_r}\right)^2
\nonumber \\&&
   -\left[ ~~ a_{41}
\frac{N( 2 + N )^2}
{324 \,(4\pi)^7}
 +
  a_{42}
\frac{N\,
     \left( 40 + 32\,N + 8\,N^2 +
       N^3 \right) }{648\,
     {(4\pi)}^7}
 +
  a_{43}
\frac{N\,     \left( 44 + 32\,N + 5\,N^2 \right)       }
{324\,           (4\pi)^7}
   \right.\nonumber \\&&\left.
 ~~~+
  a_{44}
\frac{N\,
     {\left( 2 + N \right) }^2}
     {324\,{(4\pi)}^7}
 +
  a_{45}\frac{N\,
     \left( 44 + 32\,N + 5\,N^2 \right)
       \,u^4}{324\,m_r^3\,
     {(4\pi)}^7}
\right]
\left(\frac{u}{m_r}\right)^ 3
+\dots.
\label{@eq9}\end{eqnarray}
 where
$a_2\equiv
\log (4/3)/2\approx0.143841 $
and the other constants are only known numerically
\cite{braatexp}:
\begin{eqnarray} \!\!\!
\begin{tabular}{ll}
$a_3~ = 0.642144,~
a_{41} = -0.115069,~
a_{42} = 3.128107,~
a_{43} = 1.63,~
a_{44} = -0.624638,~
a_{45} = 2.39.$
 \end{tabular}
\label{@10}\end{eqnarray}
Writing the above expansion  up to the $L$th term
 as
$F_L(u)=\Sigma_{l=-1}^Lf_l(u/4\pi m_r)^l$, the
expansion coefficients for the relevant number of components
 $N=2$
are
\cite{braatexp}:
\begin{eqnarray}  \!\!\!\!\!\!\!\!
f_{-1}=
-126.651\times 10^{-4}
,~~f_0=0,~~
f_1=-4.04837 \times  10^{-4},~~
f_2=  2.39701\times {10}^{-4},~
f_3=-1.80\times {10}^{-4}.
\label{@fexpp}\end{eqnarray}
\comment{
For comparison with large-$n$
calculations we also give the expansion in this limit:
\begin{eqnarray}  \!\!\!\!\!\!\!\!
\left\langle\frac{ \Delta \phi^2}{{u}}\right\rangle=F^{n=\infty}_3
&\equiv&
-4.02719\,{10}^{-6}\,u +
  1.90442\,{10}^{-8}\,u^2 -
  9.85172\,{10}^{-11}\,u^3   \nonumber \\
&+&
  \left( -8.05437\,{10}^{-6}\,
      u + 1.90442\,{10}^{-7}\,u^2 -
     2.27976\,{10}^{-9}\,u^3 \right)\frac{1}{n}
\nonumber \\&
+&   \left( 3.04707\,{10}^{-7}\,u^2 -
     1.2443\,{10}^{-8}\,u^3 \right)  \frac{1}{n^2}
.
\label{@}\end{eqnarray}
}
We need the
value of the series
$F_L(u)$ in the critical limit $m_r\rightarrow 0$,
which is obviously equivalent
to the
strong-coupling limit of $F_L(u)$.
As mentioned above, this limit should be
most accurately found
with the help of
variational perturbation theory \cite{SC3,SCE,KS}.

If the series were of quantum mechanical
origin,
we could find this limit
by applying the
rules of {\em naive\/} variational perturbation theory
\cite{PI}.
We form the sequence of truncated expansions
$F_L(u)$ for $1,2,3$
 and replace each
term
\begin{equation}
(u/m_r)^l\rightarrow  K^l[1-1]^{-l/2}_ {L-l}
\label{@rep}\end{equation}
where the symbol
$[1-1]^{r}_ {k}$
is defined as the
 binomial expansion of $(1-1)^r$ truncated
after  the
$k$th term
\begin{equation}
[1-1]^{r}_ {k}\equiv \sum_{i=0}^{k}\left( r \atop i\right)(-1)^i=(-1)^k
\left( r -1\atop k\right).
\label{@}\end{equation}
The resulting  expressions
must be optimized in the variational
parameter $K$.
They are listed
 in Table \ref{vexp}.
\begin{table}[tbhp]
\caption[]{Trial functions
for the naive quantum-mechanical variational perturbation expansion
 }
\begin{tabular}{lll}
&$W^{\rm QM}_1={-0.0596831}{K^{-1}} - 0.0000322159\,K,$\\
&$W^{\rm QM}_2={-0.0497359}{K^{-1}} - 0.0000483239\,K +
  1.51792\,{10}^{-6}\,K^2,$\\
&$W^{\rm QM}_3={-0.0435189}{K^{-1}} - 0.0000604049\,K +
  3.03584\,{10}^{-6}\,K^2 -
  .908\,{10}^{-7}\,K^3.$
\end{tabular}
\label{vexp}\end{table}
The approximants
$W^{\rm QM}_{1,2,3}$
have  extrema
$W^{\rm QM ext}_{1,2,3}
\approx -0.00277,\,+0.00405,\, -0.0029,$
corresponding, via (\ref{@NumF}), to $c_1\approx3.059,\,-4.46,\,3.01$.
These values have previously been obtained in Ref.~\cite{ramospr}
in a much more complicated way via a so-called $ \delta $-expansion.
Note the negative sign of the second approximation
arising from the fact that an extremum exists only at negative $K$.
According to our rules of variational perturbation theory
one should, in this case,
 use the saddle point at positive $K$
which would yield
 $W_2^{\rm QM}=
-0.00153$
 corresponding to
$c_1\approx1.69$ rather than -4.46, leading to the
more reasonable
 approximation sequence
$c_1\approx3.059,\,1.69,\,3.01$, which shows no sign of convergence.
In
$W^{\rm QM}_{3}$,
there is also a pair of complex extrema
from which the authors of Ref.~\cite{ramospr}
extract the real part Re $\tilde W
_{3\,\rm complex}
^{\rm QM}
\approx-0.00134$
 corresponding to
$c_1\approx1.48$,
which they state as
 their
final result.
There is, however, no acceptable
theoretical justification for such a choice
\cite{Ham}.

This lack of convergence is not astonishing
since  naive
quantum mechanical
variational perturbation theory is
inapplicable to field theory,
contrary to
ubiquitous statements in the literature \cite{ubi}.
A simple but essential  modification is necessary
to allow for the well-known fact that there
are {\em anomalous dimensions\/}
in the critical regime of fluctuating fields.
This modification
was discovered
in Ref.~\cite{SC3}
and tested by the fact it reproduces
in $D=4- \epsilon $ dimensions
{\em exactly\/}
the known $ \epsilon $ expansions
of renormalization group theory
\cite{SCE}.
In $D=3$ dimensions, it leads
 to the most accurate critical exponents so far
(see in particular Chapters 19 and 20 in the textbook \cite{KS}).
\comment{By expanding the results of variational perturbation
theory in $4- \epsilon $ dimensions
in powers of $ \epsilon $, the field-theoretic
variational perturbation
theory
 reproduces precisely the
$ \epsilon $-expansions of renormalization group theory \cite{SCE}.
}

The correct procedure goes as follows:
We form the logarithmic derivative
of the expansion (\ref{@eq9}):
\begin{eqnarray}
 \beta\left({u}\right) \equiv \frac{\partial \log F(u)}{\partial \log u}
=-1 +
2\frac{f_1}{f_{-1}}\left(\frac{u}{m_r}\right)^2
+3\frac{f_2}{f_{-1}}
\left(\frac{u}{m_r}\right)^3
+\left(
4\frac{f_3}{f_{-1}}
-2\frac{f_1^2}{f^2_{-1}} \right)
\left(\frac{u}{m_r}\right)^4+\dots~.
 \label{@betaf}\end{eqnarray}
In order for $F(u)$ to go to a constant
in the critical limit
$m_r\rightarrow 0$,
this function must go to zero
in the strong-coupling limit
$u\rightarrow \infty$.
Writing the expansion
as
$\beta _L
\left({u}\right)= -1+
\Sigma
_{l=2}^L\, b_l(u/4\pi m_r)^l
$, the
coefficients are
\begin{equation}
 b _2=0.0639293,~~~
 b_3=-0.056778,~~~
 b _4=0.0548799.
\label{@}\end{equation}
The sums
$ \beta_L(u)$ have to be evaluated
for
 $u \rightarrow \infty $
allowing for the universal anomalous dimension $ \omega $
by which the
physical observables
of $\phi^4$-theories approach the scaling limit \cite{zj,KS}.
The approach to the critical point  $A+B(m_r/u)^{ \omega '}$
where $ \omega '=
 \omega /(1-\eta /2)$ \cite{rem}.
The exponent
 $ \eta $ is the small anomalous dimension of the field
 while $ \omega $ is related to the
 famous Wegner exponent
\cite{Wegner}
of
renormalization group theory $ \Delta \equiv  \omega  \nu $.
Here it
appears
in the variational
expression for the
 strong-coupling limit
which is found  \cite{SC3,SCE} by replacing
$(u/m_r)^l$ by $ K^l[1-1]^{-ql/2}_ {L-l} $,
where $q\equiv 2/ \omega' $.
Thus we obtain the
variational expressions
\begin{eqnarray}
W^ \beta _3&=&
-1 +
\left( \frac{2\,{f_1}}
      {{f_{-1}}}
+
     \frac{2\,{f_1}\,q}
      {{f_{-1}}} \right)
  K^2
  +\frac{3\,{f_2}
}
   {{f_{-1}}}
K^3
\\
W^ \beta _4&=&-1+
\left( \frac{2\,{f_1}}
       {{f_{-1}}}
 +
      \frac{3\,{f_1}\,q}
       {{f_{-1}}} +
      \frac{{f_1}\,q^2}
       {{f_{-1}}} \right)
   K^2
  +
\left( \frac{3\,{f_2}}
       {{f_{-1}}} +
      \frac{9\,{f_2}\,q}
       {2\,{f_{-1}}} \right)
  K^3
 +
   \left( \frac{-2\,{{f_1}}^2}
       {{{f_{-1}}}^2} +
      \frac{4\,{f_3}}
       {{f_{-1}}} \right) \,K^4
  \label{@}\end{eqnarray}
The first has a vanishing extremum at
$ \omega' _3=0.592$, the second has
neither an extremum nor a saddle point.
However, a complex pair of extrema
lies reasonably close to the real axis at $ \omega' _4=0.635\pm0.116$,
 whose real part is
not far from
 the true
 exponent of approach  $ \omega'_\infty \approx0.81$ \cite{zj,KS},
to which $ \omega' _L$  will converge for order  $L\rightarrow \infty$
 \cite{SC3}.
Given these $ \omega' $-values,
we now form
the variational expressions
 $W_L$
from  $F_L$ by the replacement
$(u/m_r)^l\rightarrow   K^l[1-1]^{-ql/2}_ {L-l} $,
which are
\begin{eqnarray}
\hspace{-5mm}
W_2&=&   f_{-1}\left(1-\frac{3}{4}q+\frac{1}{8}q^2\right)K^{-1}+f_1 K,\\
\hspace{-5mm}
W_3&=&   f_{-1}\left(1-\frac{11}{13}q+\frac{1}{4}q^2-\frac{1}{48}q^3\right)K^{-1}
+f_1\left(1+\frac{q}{2}\right) K+f_2K^2,~~~\label{@18}\\
\hspace{-5mm}
W_4&=&   f_{-1}\left(1-\frac{25}{24}q+\frac{35}{96}q^2
-\frac{5}{96}q^3
+\frac{1}{384}q^4
\right)K^{-1}
+f_1\left(1+\frac{3}{4}q+\frac{1}{8}q^2\right) K+
f_2(1+q)K^2
+f_3K^3.~~~\label{@19}
\label{@}\end{eqnarray}

\begin{figure}[bht]
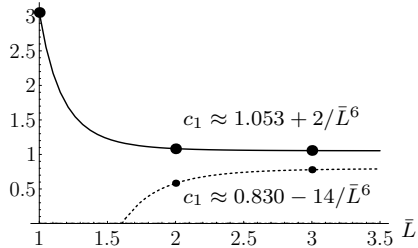

\vspace{2cm}
\input plapprno.tps
\caption[]{The three approximants for $c_1$
plotted against the order of variational
approximation  $\bar L\equiv L-1=1,2,3$,
and  extrapolation to the infinite-order limit.}
\label{figure}\end{figure}

The lowest function $W_2$ is optimized with the  naive
growth parameter $q=1$
since to this order
no anomalous value can be determined
from the zero of the beta function (\ref{@betaf}).
 The optimal result is $W_2^ {\rm opt}=-\sqrt{\log[4/3]/6}/8\pi^2\approx-0.00277$
corresponding to
$c_1\equiv 3.06$.
The next function $W_3$ is optimized with the
above determined
$q_3=2/ \omega'_3$ and yields
$W_3^ {\rm opt}\approx-0.000976$
corresponding to
$c_1\equiv 1.078$.
Although
  $ \omega' _4$ is not real
we shall insert its real part
into $W_4 $ and find
$W_4^{\rm opt}\equiv -0.000957$
corresponding to
$c_1\equiv 1.057$.
The three values of $c_1$
for $\bar L\equiv L-1=1,2,3$ can well be fitted
by a function
$c_{1}\approx1.053+2/\bar L^6$
(see Fig.~\ref{figure}).
Such a fit is suggested
by the general large-$L$ behavior
$a+be^{-c\,\bar L^{1- \omega' }}$ which was
derived in
Refs.~\cite{PI}.
Due to the smallness
of $1- \omega' \approx0.2$, this can be replaced by
$\approx a'+b'/\bar L^{s}$.

Alternatively, we may optimize the functions $W_{1,2,3}$ using the
known precise
value  of
$q_\infty=2/ \omega' _\infty\approx2/0.81$.
Then $W_2$ turns out to have no optimum, whereas
the others yield
$W^{\rm opt}_{3,4}
\approx
-0.000554,\,
-0.000735,$
  corresponding via Eq.~(\ref{@NumF})
to $c_1=0.580,\,0.773$.
If these two values
 are fitted
by the same inverse power of $\bar L$, we find
$c_{1}\approx0.83- 14/\bar L^6$.
From the extrapolations to
infinite order we estimate
$c_{1,\infty}\approx
0.92\pm0.13$.

\comment{It is possible to derive also an estimate from below.
For this we try to optimize all three functions $W_{123}$
with the precise $q_\infty\approx2/0.2$.
The first has no optimum, while $W_{2,3}\approx-0.0005289,\,0.000706$
corresponding to $c_1\approx0.583,\,0.779$.
These values are also shown in Fig.~(\ref{figure}).
They can be fitted to reach the same large-$L$ limit by a curve
$0.82-4.66/x^{4.5}$.
}

This
 result
is to be compared with latest Monte Carlo data
which estimate
$c_1\approx1.32\pm0.02$
\cite{arnold1,russos}.
Previous theoretical
estimates are
 $c_1\approx2.90$ \cite {baymprl},
$2.33$ from a $1/N$-expansion \cite {baymN}),
$1.71$ from a next-to-leading order
in a
$1/N$-expansion
 \cite {arnold},
$3.059$ from
an inapplicable $ \delta $-expansion \cite {prb}
to three loops,
and $1.48$ from
the same $ \delta $-expansion
to five loops, with a questionable
evaluation at a complex extremum \cite {ramospr} and some
wrong expansion coefficients (see \cite{braatexp}).
Remarkably, our result lies close
to the
average between the latest
and the
first
Monte Carlo result
 $c_1\approx
0.34\pm0.03$  in Ref.~\cite{Grue}.

As a cross check of the reliability of our theory
consider the result in the limit $N\rightarrow \infty$.
Here we must  drop the first term in the expansion
(\ref{@eq9})
which vanishes at the critical point
(but would diverge
 for  $N\rightarrow \infty$ at finite $m_r$).
The remaining  expansion
coefficients
of $
\left\langle{ \phi^2}/{u}\right\rangle/N$
in powers of $Nu/4\pi m_r$
are
\begin{eqnarray}  \!\!\!\!\!\!\!\!
f_1=-6.35917 \, 10^{-4},~~~
f_2=4.7315 \,{10}^{-4},~~
f_3=-3.84146\,{10}^{-4}.
\label{@fexpp}\end{eqnarray}
Using the $N\rightarrow \infty$ limit of $ \omega' $
which is equal to $1$ implying $q=2$ in Eqs.~(\ref{@18})
and (\ref{@19}),
we obtain
 the two variational approximations
\begin{eqnarray}
W_2^\infty= -0.00127183 K+0.00047315K^2,~~
W_3^\infty=  -0.00190775K+0.00141945K^2-0.000384146K^3,
\label{@}\end{eqnarray}
whose optima yield
the approximations $c_1\approx1.886$ and $2.017$,
converging rapidly towards the
exact large-$N$ result $2.33$ of Ref.~\cite {baymN}, with a 10\% error.

Numerically,  the first two $1/N$-corrections
found from a fit to
large-$N$ results
obtained by using the known large-$N$ expression for
$ \omega' =1-8( 8/3\pi^2N)+2(104/3-9\pi^2/2)(8/3\pi^2N)^2$
\cite{omN} produce a
finite-$N$  correction
factor $(1-3.1/N+30.3/N^2+\dots)$,
to be compared with
$(1-0.527/N+\dots)$ obtained in Ref.~\cite{arnold}.

Since the large-$N$ results can only be obtained
so well
without the use of the first term
we repeat the evaluations
of the series at the physical value
$N=2$ without the first term, where the variational expressions for $f$ are
\begin{eqnarray}
W_2& =&f_1\left(1+\frac{q}2\right)K+f_2K^2,\nonumber \\
W_3&=&f_1\left(1+\frac{3}{4}q+\frac{1}{8}q^2\right)K+f_2\left(1+q\right)K^2+f_3K^3.
\label{@}\end{eqnarray}
The lowest order optimum
lies now at
 $W_2^{\rm opt}=-f_1^2(2+q)^2/16f_2^2$, yielding
 $c_1\equiv 0.942$
for the exact
$q=2/0.81$.
To next order,
an optimal turning point of $W_3$
yields
 $c_1\approx 1.038$.

At this order,
we can
 derive
a
variational
expression for the determination of $ \omega '$
using the analog of
Eq.~(\ref{@betaf}) which reads
\begin{eqnarray}
 \beta\left({u}\right) \equiv \frac{\partial \log F(u)}{\partial \log u}
=1 +
\frac{f_2}{f_{1}}\,\frac{u}{m_r}
+\left(2
\frac{f_3}{f_{1}}
-\frac{f_2^2}{f_{1}^2}
\right)
\left(\frac{u}{m_r}\right)^2
+\dots~.
 \label{@betafn}\end{eqnarray}
  After the replacement (\ref{@rep})
we find
\begin{eqnarray}
W^ \beta _3&=&1+
\frac{{f_2}(1+q/2)}
       {{f_{1}}}K
 +
\left(
      2\frac{{f_3}}
       {{f_{1}}}
      -\frac{\,{f_2^2}}
       {{f_{1}^2}}
\right)
   K^2
 +\dots
  \label{@}\end{eqnarray}
whose vanishing extremum determines $ \omega '=2/q $ as being
\begin{equation}
 \omega '_3=\left(2 \sqrt{2f_1f_3/f_2^2-1}-1\right)^{-1}\approx0.675,
\label{@}\end{equation}
leading to
$c_1\approx1.238$
from an optimal turning point of $W_3$.
There are now too few points to perform  an extrapolation
to infinite order.
From the average of the two highest-order results
we obtain our
final estimate: $c_1\approx 1.14\pm0.11$,
 such that
 the critical temperature shift is
\begin{equation}
\frac{ \Delta T_c}{T_c^{(0)}}
\approx
(1.14\pm0.11)\,
an^{1/3}.
\label{@res}\end{equation}
This lies reasonably close to
 the Monte Carlo number $c_1\approx1.32\pm0.02$.
~\\
~\\
Acknowledgment:\\
The author acknowledges
 useful communications with
M. Pinto and R. Ramos
on their work,
as well as
F. Nogueira
and
A. Pelster
for many discussions.
He also thanks
B. Kastening
for
proofreading the paper, for
confirming the
approximate correctness of
the
Feynman integrals
of Ref.~\cite{braat}
leading to the coefficients
(\ref{@10})
rather than those of Ref.~\cite{ramospr}
(which are listed
in
 \cite{braatexp} for comparison).

~\\
~\\
Note added in proof:\\
Since this paper appeared on the Los Alamos server last October,
the Feynman diagrams where recalculated more accurately
\begin{figure}[bht]
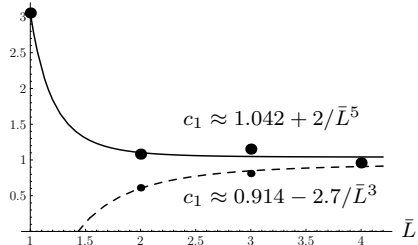

\vspace{2cm}
\input plapprnk.tps
\caption[]{The four approximants for $c_1$
plotted against the order of variational
approximation  $\bar L\equiv L-1=1,2,3,4$,
and  extrapolation to the infinite-order limit.
The second-highest aproximations are $c_1=1.115$ and $0.797$
for self-consistent  $ \omega '=0.745$
and exact $ \omega'\approx0.81 $,
the highest aproximations are $c_1=0.959$ and $0.886$
for self-consistent  $ \omega '=0.745$
and exact $ \omega'\approx0.81 $, respectively.
The latter dots cannot be distinguished.
}
\label{figuren}\end{figure}
and extended to six loops in Ref.~\cite{Kas}.
 The more accurate $f_3$ has increased by about $6\%$ to $-1.92\times 10^{-4}$,
and the new expansion coefficient is
$f_4=1.873\times 10^{-4}$.
Using the new numbers to perform a variational evaluation
 fits of the first type including the
 anomalous first term
$f_{-1}$ we find the curves
 shown in
 Fig.~\ref{figure} yielding
$c_1\approx 0.98\pm0.06$,
 only slighly larger  than
than the previous five-loop number $0.92\pm0.13$. The upper curve
is now associated
with the self-consistent $ \omega' $-values
$ \omega' _3=0.626$ and $ \omega' _4=0.745$, the last being
 almost equal to the exact value $0.81$.

The evaluation of the second type where
the anomalous term is omitted
yields now
$c_1=0.942,\,1.056,\,1.120$,
for $ \omega'=0.81$,
 and
$c_1=1.409,\,1.384$
 for the
variationally determined
$ \omega '_3=0.621$ and $0.638$, as shown in
in Ref.~\cite{Kas}.
Combining the last two results
one obtains the larger value
$c_1\equiv 1.23\pm0.12$, lying
closer
to
 the Monte Carlo number $c_1\approx1.32\pm0.02$
than our five-loop result (\ref{@res}).


\begin{thebibliography}{99}
\bibitem{braatexp}
These are the results of \cite{braat}.
They differ from those of
 Refs.~\cite{ramospr} which are
$a_3\! =\! 0.644519,
a_{41} \!=\! 0.87339,
a_{42}\! = \!3.15905,
a_{43} \!= \!1.70959,
a_{44} \!= \!4.4411,
a_{45} \!= \!2.37741.$
The coefficients of the series
(\ref{@fexpp})
to be resummed
differ mainly in the last term:
$f_{-1}=
-126.651\,10^{-4}
,f_0=0,
f_1\!=-4.04857 \, 10^{-4},
f_2\!=  2.40587\,{10}^{-4},
f_3\!=-2.06849\,{10}^{-4}.$
\bibitem{rem}
With
standard  normalization conditions
used in the 3-dimensional $\phi^4$-theory, the approach to scaling is governed  by
Wegner's exponent
$ \omega $
 (see \cite{SC3}).
The present definition of $m_r$ differs from the
the inverse correlation length $m=\xi^{-1}$ by
a factor: $m_r=mZ_\phi^{-1}\propto m\,m^{-\eta/2}$ for $m\rightarrow 0$.
This changes the exponent of approach to $ \omega '= \omega /(1- \eta /2)$.
I thank B. Kastening for
noting this.
\end{thebibliography}
\end{document}